\documentclass[preprint]{aastex}

\begin{document}

\title {The Triply Eclipsing Hierarchical Triple Star KIC002856960}
\author{Jae Woo Lee, Seung-Lee Kim, Chung-Uk Lee, Byeong-Cheol Lee, Byeong-Gon Park, and Tobias Cornelius Hinse}
\affil{Korea Astronomy and Space Science Institute, Daejon 305-348, Korea}
\email{jwlee@kasi.re.kr, slkim@kasi.re.kr, leecu@kasi.re.kr, bclee@kasi.re.kr, bgpark@kasi.re.kr, tchinse@gmail.com}

\begin{abstract}
In a recent study, Armstrong et al. presented an eclipsing binary star of about 6.2 h period with transit-like tertiary signals
occurring every 204.2 d in the {\it Kepler} public data of KIC002856960 and proposed three possible hierarchical structures: 
(AB)b, (AB)C, and A(BC). We analyzed the {\it Kepler} light curve by including a third light source and one starspot on 
each binary component. The results represent that the close eclipsing pair is in a low-mass eccentric-orbit, detached configuration. 
Based on 123 eclipse timings calculated from the Wilson-Devinney binary model, a period study of the close binary reveals that 
the orbital period has experienced a sinusoidal variation with a period and a semi-amplitude of 205$\pm$2 d and 0.0021$\pm$0.0002 d, 
respectively. The period variation would be produced by the light-travel-time effect due to a gravitationally-bound third body 
with a minimum mass of $M_3 \sin i_3$=0.76 M$_\odot$ in an eccentric orbit of $e_3$=0.61. This is consistent with the presence of 
third light found in our light-curve solution and the tertiary signal of 204.2 d period most likely arises from 
the K-type star crossed by the close eclipsing binary. Then, KIC002856960 is a triply eclipsing hierarchical system, A(BC), 
consisting of a close binary with two M-type dwarfs and a more massive K-type component. The presence of the third star may have 
played an important role in the formation and evolution of the close pair, which may ultimately evolve into a contact system 
by angular momentum loss.
\end{abstract}

\keywords{binaries: close --- binaries: eclipsing --- stars: individual (KIC002856960) --- stars: spots}{}

\section{INTRODUCTION}

Multiple star systems with eclipsing features are very rare and interesting objects for understanding the star formation and 
stellar dynamical evolution. To our knowledge, four systems have been found to be a double eclipsing binary (ADS 9537AB, 
Batten \& Hardie 1965; V994 Her, Lee et al. 2008a; KIC 4247791, Lehmann et al. 2012; CzeV343, Caga\u s \& Pejcha 2012) and 
two systems to be a triply eclipsing hierarchical triple (HD 181068, Derekas et al. 2011; KOI-126, Carter et al. 2011). 
In the case of hierarchical triple star systems (double+single), outer third components may remove angular momentum from 
the inner close pairs via Kozai oscillation (Kozai 1962; Pribulla \& Rucinski 2006) or a combination of the Kozai cycle and 
tidal friction (Fabrycky \& Tremaine 2007). Kozai cycles periodically raise the eccentricity of the close binary while 
the tidal friction efficiently dissipates the orbital energy during the close pericenter passages. As a result, 
the inner close binary undergoes an orbital period decrease and forms the tidal-locked detached binary with 
a short orbital period. 

The presence of a third body orbiting an eclipsing binary causes a periodic variation of the eclipsing period due to 
the increasing and decreasing light-travel times (LTT) to the observer (Irwin 1952, 1959). The timing variability may be 
represented by an $O$--$C$ diagram, showing the differences between the observed ($O$) and the calculated ($C$) eclipse epochs. 
Because times of minimum light act as an accurate clock, they have been used as a tool that detects the third components 
to close binary stars. With timing accuracies of about $\pm$10 s, it should be possible to detect circumbinary planets 
of $\sim$10 M$\rm_{Jup}$ in longer periods around eclipsing systems (Ribas 2006; Lee et al. 2009; Pribulla et al. 2012).

In a recent work, Armstrong et al. (2012) reported a very interesting discovery of an eclipsing binary star with 
tertiary signals in the {\it Kepler} public data of KIC002856960 (RA$_{2000}$=19$^{\rm h}$29$^{\rm m}$31$\fs52$; 
DEC$_{2000}$=+38$^{\circ}$04${\rm '}$35$\farcs$9; $K_{\rm p}$=$+$15.615;  $T_{\rm eff}$=4733 K). The close binary components 
display eclipses with a 6.2 h (0.2585 d) period, while the outer third body leads to transit-like features in the light curve 
occurring every 204.2 days. They suggested that the tertial signals may result from either a planetary companion 
(or low mass dwarf star) orbiting around a close eclipsing binary, (AB)b/C, or a close binary orbiting around a more massive star, 
A(BC) (named C(AB) in their paper), but were unable to distinguish between them. In this paper, we analyze in detail the light curve 
and eclipse timings from the {\it Kepler} photometric data, and show that KIC002856960 has evidences of a triple star system, 
consisting a close binary and a more massive third companion forming a wider binary with the mass center of the close pair.

\section{{\it KEPLER} LIGHT-CURVE ANALYSIS}

For our study, we used 22,695 individual observations (BJD 2454964$-$2455462) from the first six quarters of {\it Kepler} operations.
Detailed information about the {\it Kepler} spacecraft and its performance can be found in Koch et al. (2010). The light curve of 
KIC002856960 is plotted in Figure 1, as normalized flux {\it versus} orbital phase, which was computed according to the linear terms 
of the LTT ephemeris described in the following section. In order to determine the photometric solutions of the binary pair, 
the {\it Kepler} data was analyzed by using the 2003 version of the Wilson-Devinney synthesis code (Wilson \& Devinney 1971, hereafter W-D). 
For this purpose, the mean light level at phase 0.70 was set to unity. The logarithmic bolometric ($X$, $Y$) and monochromatic 
($x$, $y$) limb-darkening coefficient were interpolated from the values of van Hamme (1993) in concert with 
the model atmosphere option. The gravity-darkening exponents and bolometric albedoes were held fixed at standard values 
($g$=0.32 and $A$=0.5) for stars with convective envelopes. A mass ratio of $q$=1.0 and a circular orbit ($e$=0.0) 
were initially taken from visual inspection of the light curve and a synchronous rotation for both components was adopted. 
Furthermore, a third light source ($\ell_3$) was considered. In Table 1, the parameters with parenthesized formal errors signify 
the adjusted ones and the subscripts 1 and 2 refer to the primary and secondary stars being eclipsed at Min I and Min II, respectively. 

The effective temperature of KIC002856960 is given as 4733 K from {\it Kepler} Input Catalogue (KIC, Brown et al. 2011), which would 
correspond to a spectral type of approximately K3. First of all, we analyzed the {\it Kepler} data by adopting this value as a temperature 
of the primary component. The preliminary analysis reveals $\ell_3$ to contribute about 97\% to the total luminosity of 
this system. This indicates that the KIC temperature may be mainly come from the third component and then that the close binary may 
be composed of two M-type components. Thus, the surface temperature of the hotter primary star was initialized to be $T_{1}$=3150 K.

Our light-curve analyses have been performed through three stages. In the first stage, the light curve was solved without spots.
The result for this unspotted model is plotted as the dashed curve in Figure 1, where the computed light curves do not fit well with 
the {\it Kepler} data. This discrepancy in low-mass close binaries can be explained by spot activity on binary components
as a magnetic dynamo effect (e.g. Lee et al. 2012). In the second stage, we analyzed the light curve by using the unspotted solution 
as the initial values and including the starspot on the component stars. The result for the spot model is given as Model 1 in Table 1 
and displayed as the solid curves in Figure 1. The light variation was satisfactorily modelled by using a two-spot model with 
one starspot on each component. In the third stage, although the mass ratio has little effect for detached systems, $q$, $e$, 
and $\omega$ were included as additional free parameters. The final result appears as Model 2 in Table 1. In a formal sense, 
the Model 2 gives a smaller value for the sum of the residuals squared, $\Sigma W(O-C)^2$, than Model 1. 
The upper panel of Figure 2 represents all light residuals from Model 2 and the lower three panels are drawn to see the details 
of three tertiary signals with several transits/eclipses. The period and duration of the tertiary signal were estimated to be 
about 204.2 d and 1.3 d and its maximum depth is 6.8\% to the system flux, which is about 6 times deeper than the close binary eclipse. 
These values are almost consistent with those given by Armstrong et al. (2012), but the light residuals from the W-D binary model 
presented in this paper may be more reasonable to resolve the tertiary signal to the {\it Kepler} data.

Assuming the  primary star to be a normal main-sequence one with a spectral type of about M6, we estimated the absolute dimensions 
for Model 1 and Model 2 from our photometric solutions and from Harmanec's (1988) relation between the spectral type 
(effective temperature) and stellar mass. These are given in the entries on the last three lines of Table 1. The radii are 
the mean volume radii calculated from the tables of Mochnacki (1984) and the luminosity ($L$) was computed by adopting 
$T_{\rm eff}$$_\odot$=5,780 K and $M_{\rm bol}$$_\odot$=+4.73 for solar values.

\section{ECLIPSE TIMING VARIATION}

Because the {\it Kepler} data were taken in long cadence mode of 29.4 min, corresponding to about 8\% of the close binary 
orbital period, it is impossible to measure a minimum epoch for each eclipse curve. Thus, we combined the {\it Kepler} data 
at the intervals of 15 orbital periods (15$\times$$P$=3.8775 d) and made total 129 light curves. Then, the times of 
minimum light for the separate datasets were calculated by means of adjusting only the ephemeris epoch ($T_0$) in the Model 2 
of Table 1, and the results are listed in Table 2. As an example, 12 datasets of them are displayed in Figure 3, where 
the eclipse timing variations are evident visually from the light curves. For the ephemeris computations, we used 
123 minimum epochs except for six timings, based on tertiary signals and insufficient data caused by gaps between quaters. 
Weights were calculated as the inverse squares of the timing errors and were then scaled from the standard deviations 
($\sigma$=0.00036 d). 

As the first step,  we applied a periodogram analysis to the timing data using the PERIOD04 program (Lenz \& Breger 2005). 
The result is plotted in the upper panel of Figure 4, which shows a peak frequency at $f$=0.00497 cycle d$^{-1}$ with 
a semi-amplitude of about 0.002 d. The observed frequency becomes a period of 201 d and can be interpreted as the LTT effect 
due to the existence of a third body. Thus, we fitted the eclipse timings to the following LTT ephemeris:
\begin{eqnarray}
C = T_0 + PE + \tau_3,
\end{eqnarray}
where $\tau_{3}$ is the LTT due to a third object in the system (Irwin 1952, 1959) and includes five parameters 
($a_{\rm b}\sin i_3$, $e_{\rm b}$, $\omega_{\rm b}$, $n_{\rm b}$, $T_{\rm b}$). Here, $a_{\rm b} \sin i_3$, $e_{\rm b}$ 
and $\omega_{\rm b}$ are the orbital parameters of the close binary around the mass center of the triple system. The parameters 
$n_{\rm b}$ and $T_{\rm b}$ denote the Keplerian mean motion of the mass center of the binary components and its epoch of 
periastron passage, respectively. The Levenberg-Marquart technique (Press et al. 1992) was applied to solve for 
the seven parameters of the ephemeris and the results are summarized in Table 3, together with related quantities. 
The $O$--$C$ residuals calculated from the linear terms in equation (1) are plotted in the upper panel of Figure 5, 
where the continuous curve represents the LTT orbit and the lower panel shows the residuals from the complete ephemeris. 
These appear as $O$--$C_{\rm full}$ in the fourth column of Table 2. In case that the LTT period is very short, 
the eclipse timing variations could be partly caused by the perturbative effect of the third component added to 
the geometrical LTT effect  (Borkovits et al. 2011, 2012). We computed the semi-amplitude of the dynamic perturbation 
on the motion of the close pair to be about 0.00003 d and found that its contribution is not significant.

The LTT orbit has a period of $P_{\rm b}$=205.2 d, a semi-amplitude of $K_{\rm b}$=0.0021 d, a projected orbital semi-major 
axis of $a_{\rm b} \sin i_3$=0.45 AU, and an eccentricity of $e_{\rm b}$=0.61. The mass function of this object becomes 
$f(M_{3})$=0.295 $M_\odot$. If the orbit of the third companion is coplanar with the close pair ($i_3$=85$^\circ.2$), 
its mass is $M_3$=0.76 M$_\odot$. We determined $R_3$$\simeq$1.27 R$_\odot$ by combining the third-body mass 
from the timing study with the third light ($\ell_3$=97\%) detected in our light-curve solution, where there is 
independent evidence to support the LTT hypothesis. This value is about 1.6 times larger than the radius of 0.79 R$_\odot$ 
obtained by means of using an empirical mass-radius relation for main-sequence stars (Southworth 2009). As displayed in 
the lower panel of Figure 4, the PERIOD04 program was used again to look for another modulation in the LTT residuals, 
but we found no credible periodicity.

\section{DISCUSSION AND CONCLUSIONS}

Our results from both the light curve and the eclipsing timings represent that KIC002856960 is a triple star system, A(BC), 
which comprises a low-mass detached close binary (BC) with a period of 0.2585 d and a more massive star (A) with an LTT period 
of about 205 d. The tertiary signals of 204.2 d period detected by Armstrong et al. (2012) and confirmed in this paper most 
likely arise from a K-type third star crossed by the close eclipsing binary with two M-type dwarfs. Then, KIC002856960 
is a third member of the triply eclipsing hierarchical systems and contains the most compact close binary with the period ratio 
of $P_{\rm A(BC)}$/$P_{\rm BC}\simeq$790. As in the cases of KOI-126 (Carter et al. 2011) and HD 181068 (Derekas et al. 2011;
Borkovits et al. 2012), the wide, single star of this system is the most massive and luminous component. The close binary stars 
in KIC002856960 are thought to have masses low enough to become fully convective and radii larger than predicted by stellar models, 
where the secondary component is oversize by about 60\%. The discrepancy may be mainly caused by the effect of magnetic activity 
in the close binary pair with both a short period and a deep convective envelope (Feiden \& Chaboyer 2012; Torres 2012).

The primary eclipse times of the close binary calculated from the LTT ephemeris are given as the dashed vertical lines in 
the lower panels of Figure 2. We can see that the deeper observed transits lie between the predicted timings. 
In the tertiary signals, the close binary passes across the disk of the K-type star. Then, the eclipsing pair partially 
blocks the surface of the larger and more massive star, so the brightness of the triple system becomes slightly faint. 
The light variation depth is proportional to the blocking area. Thus, the deeper transits may be produced when both binary components 
is displayed to observer, rather than when one of the two is in front of the K-type star. On the other hand, the mid-eclipse times 
(hereafter tertiary secondary eclipse) of the close pair eclipsed by the K-type star were estimated to be BJD 2455020.886, 
2455225.101, and 245429.316 using the parameters ($e_{3}$=0.61 and $\omega_{3}$=173 deg) of the LTT orbit. 
We looked for a possible eclipse/transit around the computed times but detected no signals. This may be a result of the fact 
that the tertiary secondary eclipse is much shallower than its primary eclipse (the K-type star crossed by the close pair). 

In close binaries that have tidal forces strong enough to cause synchronization of components, the orbital angular momentum 
is tidally coupled to the spin angular momentum. In order for the spin-orbit coupling to work efficiently, 
the initial orbital periods should be short (P $\la$ 5.0 d; Bradstreet \& Guinan; Pribulla \& Rucinski 2006; Lee et al. 2008b). 
Star formation in a triple system may alleviate the close-binary formation difficulty by re-distributing most of 
the angular momentum of a close binary to the more distant component and by leaving a low angular momentum remnant with 
a short initial orbital period. The more massive third component in KIC002856960 may have played an important role 
in the formation and evolution of the close pair, which would cause it to evolve into a contact configuration by 
angular momentum loss via magnetic braking and ultimately to coalesce into a single rapid-rotating star. 

The triply eclipsing nature would clearly make KIC002856960 an ideal target for dynamical evolutionary studies and for testing 
tidal friction theories in hierarchical triple systems. Future high-resolution spectroscopy and follow-up photometry of 
tertiary signals will help reveal more accurate properties such as the absolute dimensions and evolutionary status of the triple system.

\acknowledgments{ }
This paper includes data collected by the {\it Kepler} mission. {\it Kepler} was selected as the 10th mission of the Discovery Program. 
Funding for the {\it Kepler} mission is provided by the NASA Science Mission directorate. We have used the Simbad database maintained 
at CDS, Strasbourg, France. This work was supported by the KASI (Korea Astronomy and Space Science Institute) grant 2012-1-410-02.
T.C.H. acknowledges financial support from the Korea Research Council for Fundamental Science and Technology (KRCF) through 
the Young Research Scientist Fellowship Program.

\clearpage
\begin{figure}
 \includegraphics[]{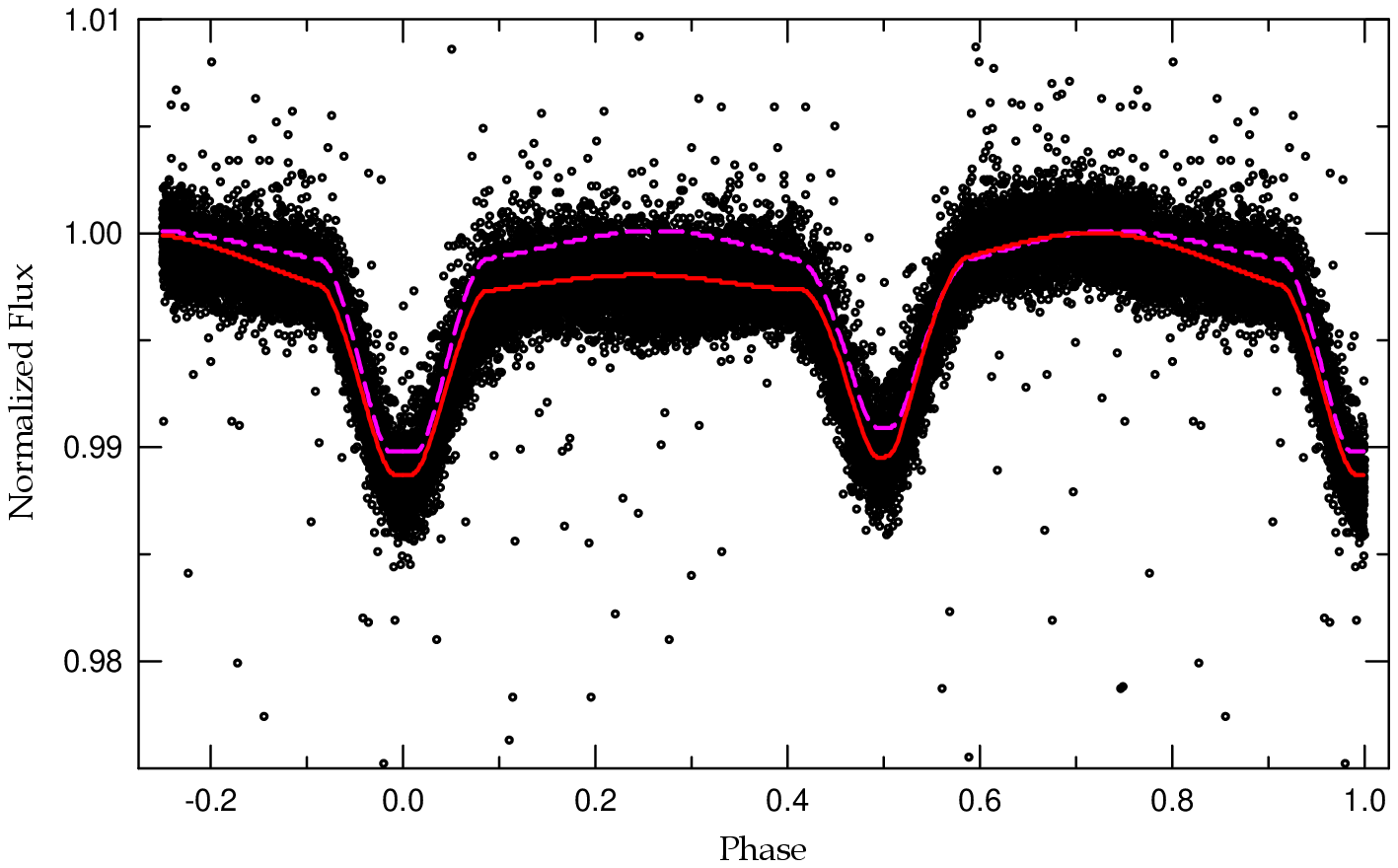}
 \caption{Light curve of KIC002856960 with the fitted models. The circles are individual measures from the {\it Kepler} spacecraft 
 and the dashed and solid lines represent the synthetic curves obtained from no spot and the two-spot model, respectively. 
 Fifty-three data points out of the flux range between 0.975 and 1.010 are not plotted to see in detail the eclipsing light curve. }
 \label{Fig1}
\end{figure}

\begin{figure}
 \includegraphics[]{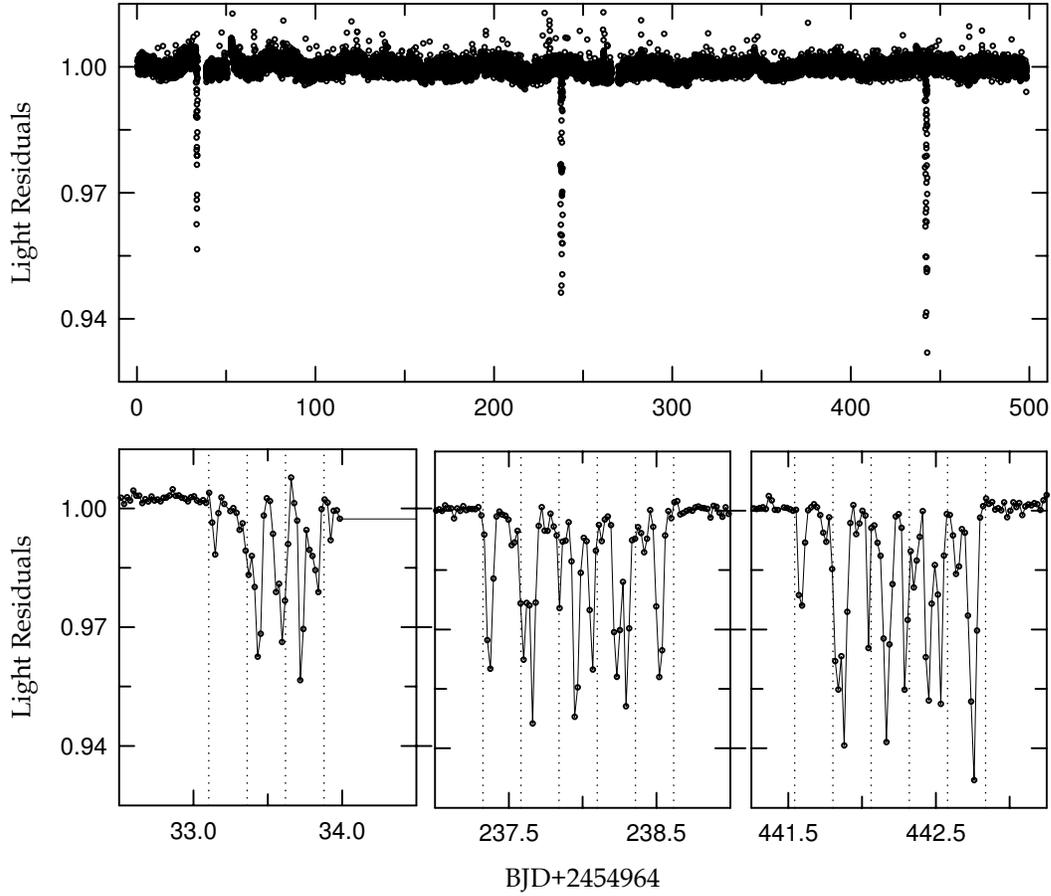}
 \caption{Light residuals corresponding to Model 2 listed in Table 1. The tertiary signals are plotted in the lower three panels,
 wherein the vertical dotted lines indicate the primary eclipse times of the close pair predicted from our LTT ephemeris. }
 \label{Fig2}
\end{figure}

\begin{figure}
 \includegraphics[]{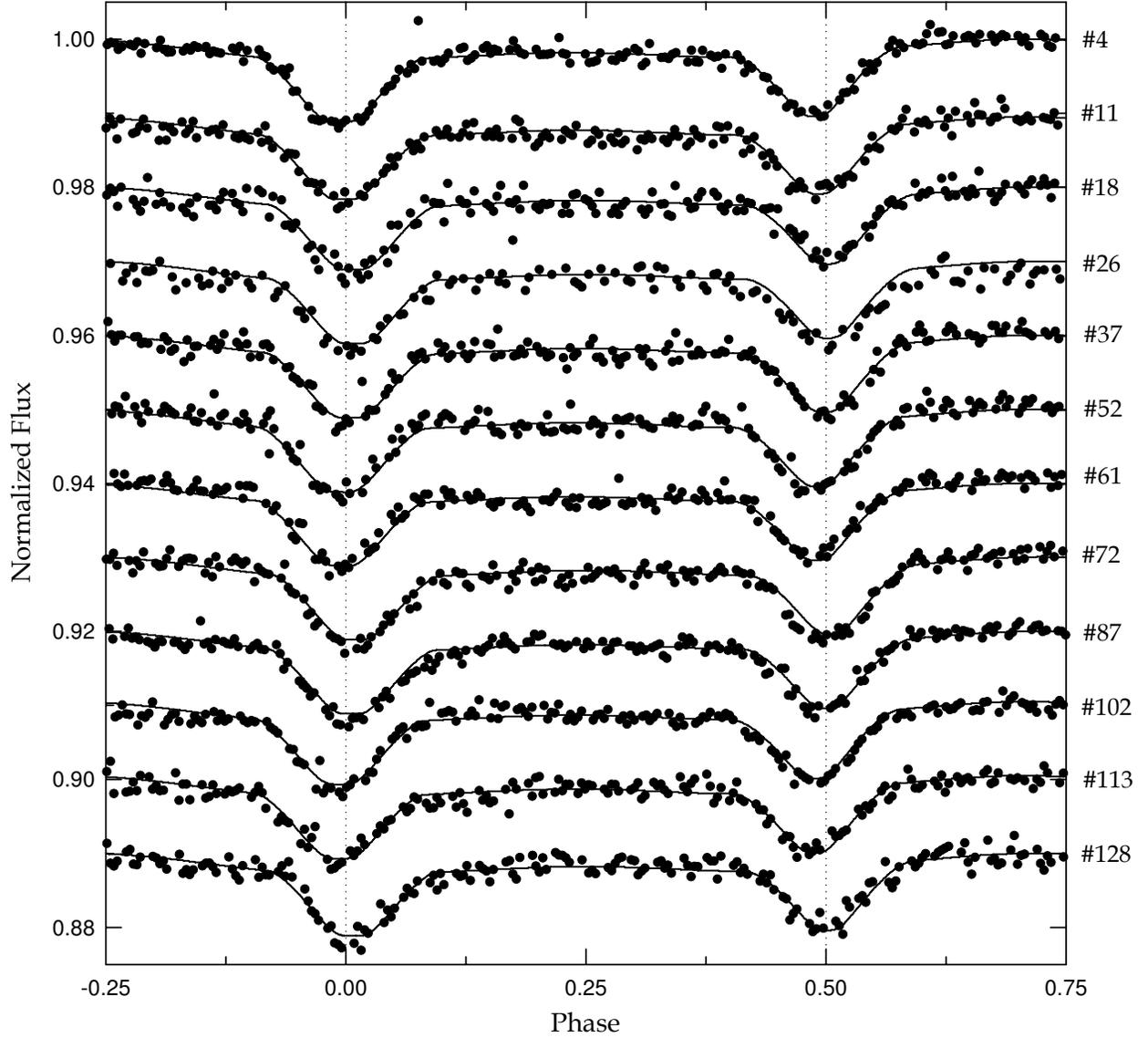}
 \caption{{\it Kepler} data of KIC002856960 combined at the intervals of 15 orbital periods. The eclipse timing variations are evident 
 visually from the light curves offset from 1.0. }
 \label{Fig3}
\end{figure}

\begin{figure}
 \includegraphics[]{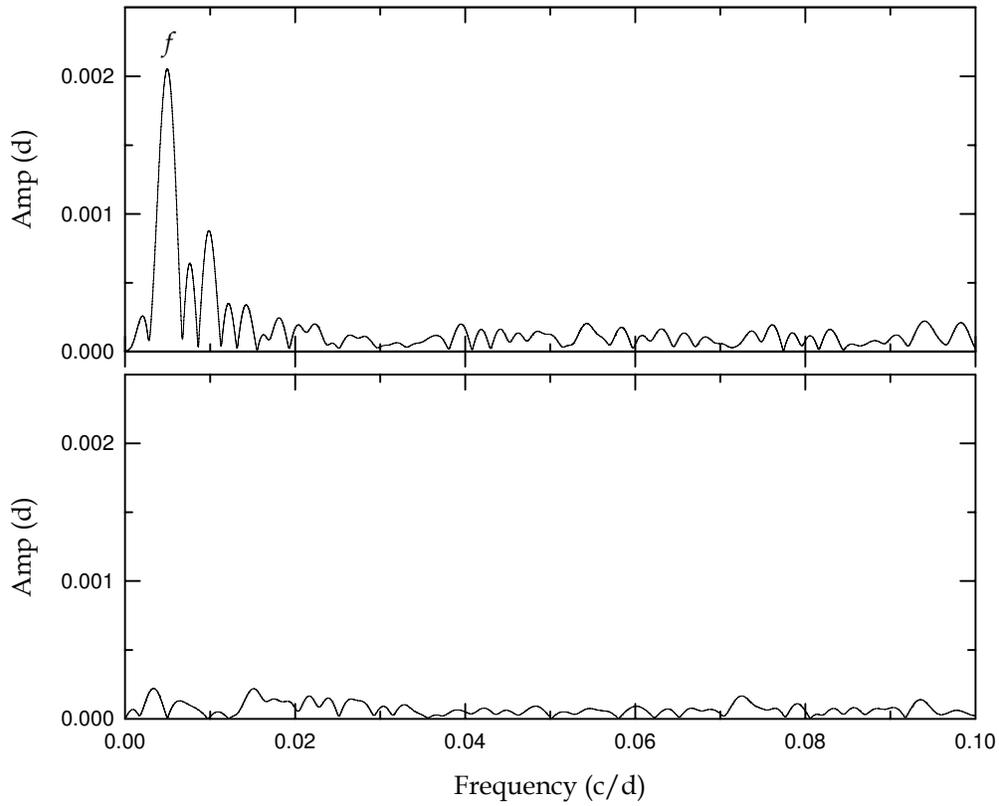}
 \caption{Periodogram from the PERIOD04 program (Lenz \& Breger 2005) for eclipse timings. A dominant frequency of 
 $f$=0.00497 cycle d$^{-1}$ is detected with a semi-amplitude of 0.00205 d and this becomes a period of 201 d.
 The amplitude spectrum for the LTT residuals is displayed in the lower panel. }
 \label{Fig4}
\end{figure}
 
\begin{figure} 
 \includegraphics[]{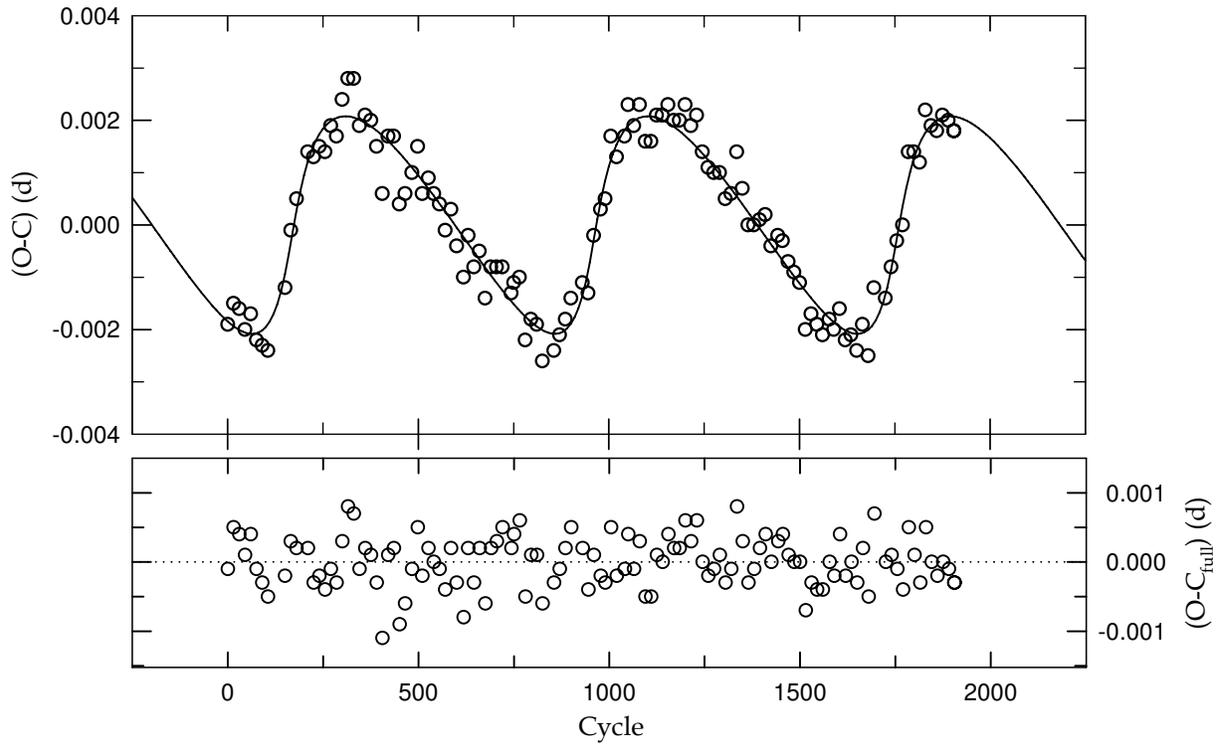}
 \caption{$O$--$C$ diagram for the close binary of KIC002856960. In the upper panel, constructed with the linear terms of Table 3, 
 the continuous curve represents the LTT orbit. The residuals from this LTT ephemeris are plotted in the lower panel.}
 \label{Fig5}
\end{figure}

\clearpage
\begin{deluxetable}{lccccc}
\tabletypesize{\small} %{\scriptsize}
\tablewidth{0pt} 
\tablecaption{Close binary parameters of KIC002856960.}
\tablehead{
\colhead{Parameter}                      & \multicolumn{2}{c}{Model 1}                 && \multicolumn{2}{c}{Model 2}                 \\ [1.0mm] \cline{2-3} \cline{5-6} \\[-2.0ex]
                                         & \colhead{Primary} & \colhead{Secondary}     && \colhead{Primary} & \colhead{Secondary}         
}
\startdata                                                                                                                                
$q$                                      & \multicolumn{2}{c}{1.0}                     && \multicolumn{2}{c}{1.106(82)}               \\
$e_{\rm b}$                              & \multicolumn{2}{c}{0.0}                     && \multicolumn{2}{c}{0.0064(14)}              \\
$\omega_{\rm b}$ (deg)                   & \multicolumn{2}{c}{}                        && \multicolumn{2}{c}{161(26)}                 \\
$i$ (deg)                                & \multicolumn{2}{c}{85.32(88)}               && \multicolumn{2}{c}{85.17(87)}               \\
$T$ (K)                                  & 3,153(160)        & 3,054(153)              && 3,160(150)        & 3,067(142)              \\
$\Omega$                                 & 5.93(11)          & 4.28(2)                 && 5.87(11)          & 4.51(20)                \\
$\Omega_{\rm in}$                        & \multicolumn{2}{c}{3.75}                    && \multicolumn{2}{c}{3.92}                    \\
$A$                                      & \multicolumn{2}{c}{0.5}                     && \multicolumn{2}{c}{0.5}                     \\
$g$                                      & \multicolumn{2}{c}{0.32}                    && \multicolumn{2}{c}{0.32}                    \\
$X$, $Y$                                 & 0.485, 0.276      & 0.485, 0.276            && 0.485, 0.276      & 0.485, 0.276            \\
$x$, $y$                                 & 0.748, 0.255      & 0.748, 0.255            && 0.748, 0.255      & 0.748, 0.255            \\
$l$/($l_{1}$+$l_{2}$+$l_{3}$)            & 0.010(1)          & 0.019                   && 0.010(1)          & 0.018                   \\
{\it $l_{3}$$\rm ^a$}                    &  \multicolumn{2}{c}{0.972(1)}               && \multicolumn{2}{c}{0.972(1)}                \\
$r$ (pole)                               & 0.2021(43)        & 0.3009(22)              && 0.2093(60)        & 0.3071(173)             \\
$r$ (point)                              & 0.2070(47)        & 0.3330(34)              && 0.2157(71)        & 0.3397(286)             \\
$r$ (side)                               & 0.2038(44)        & 0.3097(24)              && 0.2114(63)        & 0.3161(196)             \\
$r$ (back)                               & 0.2062(46)        & 0.3229(29)              && 0.2145(68)        & 0.3292(234)             \\
$r$ (volume)$\rm ^b$                     & 0.2041            & 0.3115                  && 0.2118            & 0.3179                  \\ [1.0mm]
\multicolumn{6}{l}{Spot parameters:}                                                                                                  \\        
Colatitude (deg)                         & 39.2(2.3)         & 69.6(4.7)               && 39.2(2.6)         & 69.6(3.1)               \\        
Longitude (deg)                          & 358.4(2.2)        & 65.1(2.2)               && 2.65(1.3)         & 67.9(1.1)               \\        
Radius (deg)                             & 38.2(1.4)         & 24.1(6)                 && 38.2(1.4)         & 24.1(4)                 \\        
$T$$\rm _{spot}$/$T$$\rm _{local}$       & 0.850(11)         & 0.869(7)                && 0.853(9)          & 0.875(5)                \\
$\Sigma W(O-C)^2$                        & \multicolumn{2}{c}{0.0101}                  && \multicolumn{2}{c}{0.0090}                  \\ [1.0mm]
\multicolumn{6}{l}{Absolute parameters:}                                                                                              \\            
$M$($M_\odot$)                           & 0.22              &  0.22                   && 0.22              &  0.24                   \\
$R$($R_\odot$)                           & 0.26              &  0.40                   && 0.28              &  0.42                   \\
$L$($L_\odot$)                           & 0.006             &  0.012                  && 0.007             &  0.014                  \\
\enddata
\tablenotetext{a}{Value at 0.70 phase.}
\tablenotetext{b}{Mean volume radius.}
\end{deluxetable}

\begin{deluxetable}{lcrr}
\tablewidth{0pt}
\tablecaption{Observed eclipse timings for the close binary of KIC002856960.}
\tablehead{
\colhead{BJD}   & \colhead{Error} & \colhead{$E$} & \colhead{$O$--$C_{\rm full}$} 
}
\startdata
2,454,964.78685   & $\pm$0.00031    &    0          & $-$0.000095          \\
2,454,968.66495   & $\pm$0.00027    &   15          & $+$0.000475          \\
2,454,972.54245   & $\pm$0.00026    &   30          & $+$0.000429          \\
2,454,976.41969   & $\pm$0.00032    &   45          & $+$0.000102          \\
2,454,980.29760   & $\pm$0.00031    &   60          & $+$0.000418          \\
2,454,984.17471   & $\pm$0.00062    &   75          & $-$0.000100          \\
2,454,988.05220   & $\pm$0.00036    &   90          & $-$0.000282          \\
2,454,991.92972   & $\pm$0.00057    &  105          & $-$0.000495          \\
2,454,995.80807   & $\pm$0.00368    &  120          &                      \\
2,455,003.04867   & $\pm$0.00141    &  148          &                      \\
2,455,003.56380   & $\pm$0.00031    &  150          & $-$0.000232          \\
2,455,007.44252   & $\pm$0.00032    &  165          & $+$0.000282          \\
2,455,011.32069   & $\pm$0.00039    &  180          & $+$0.000213          \\
2,455,017.00905   & $\pm$0.00154    &  202          &                      \\
2,455,019.07687   & $\pm$0.00044    &  210          & $+$0.000210          \\
2,455,022.95433   & $\pm$0.00029    &  225          & $-$0.000251          \\
2,455,026.83218   & $\pm$0.00043    &  240          & $-$0.000236          \\
2,455,030.70975   & $\pm$0.00035    &  255          & $-$0.000437          \\
2,455,034.58780   & $\pm$0.00051    &  270          & $-$0.000106          \\
2,455,038.46528   & $\pm$0.00058    &  285          & $-$0.000306          \\
2,455,042.34357   & $\pm$0.00039    &  300          & $+$0.000338          \\
2,455,046.22161   & $\pm$0.00055    &  315          & $+$0.000758          \\
2,455,050.09918   & $\pm$0.00053    &  330          & $+$0.000731          \\
2,455,053.97596   & $\pm$0.00035    &  345          & $-$0.000068          \\
2,455,057.85377   & $\pm$0.00040    &  360          & $+$0.000180          \\
2,455,061.73124   & $\pm$0.00043    &  375          & $+$0.000102          \\
2,455,065.60838   & $\pm$0.00031    &  390          & $-$0.000293          \\
2,455,069.48510   & $\pm$0.00031    &  405          & $-$0.001098          \\
2,455,073.36378   & $\pm$0.00029    &  420          & $+$0.000066          \\
2,455,077.24146   & $\pm$0.00049    &  435          & $+$0.000238          \\
2,455,081.11779   & $\pm$0.00055    &  450          & $-$0.000932          \\
2,455,084.99559   & $\pm$0.00065    &  465          & $-$0.000626          \\
2,455,089.64915   & $\pm$0.00055    &  483          & $-$0.000051          \\
2,455,093.52721   & $\pm$0.00037    &  498          & $+$0.000527          \\
2,455,096.62844   & $\pm$0.00029    &  510          & $-$0.000226          \\
2,455,100.76484   & $\pm$0.00045    &  526          & $+$0.000201          \\
2,455,104.38364   & $\pm$0.00031    &  540          & $+$0.000027          \\
2,455,108.26103   & $\pm$0.00025    &  555          & $-$0.000052          \\
2,455,112.13818   & $\pm$0.00033    &  570          & $-$0.000370          \\
2,455,116.01618   & $\pm$0.00033    &  585          & $+$0.000165          \\
2,455,119.89316   & $\pm$0.00038    &  600          & $-$0.000320          \\
2,455,124.54562   & $\pm$0.00052    &  618          & $-$0.000817          \\
2,455,127.64860   & $\pm$0.00039    &  630          & $+$0.000192          \\
2,455,131.52561   & $\pm$0.00030    &  645          & $-$0.000263          \\
2,455,135.40349   & $\pm$0.00040    &  660          & $+$0.000152          \\
2,455,139.28022   & $\pm$0.00039    &  675          & $-$0.000586          \\
2,455,143.15848   & $\pm$0.00028    &  690          & $+$0.000204          \\
2,455,147.03603   & $\pm$0.00033    &  705          & $+$0.000281          \\
2,455,150.91372   & $\pm$0.00033    &  720          & $+$0.000494          \\
2,455,156.85890   & $\pm$0.00063    &  743          & $+$0.000199          \\
2,455,158.66863   & $\pm$0.00063    &  750          & $+$0.000433          \\
2,455,162.54633   & $\pm$0.00035    &  765          & $+$0.000637          \\
2,455,166.42271   & $\pm$0.00032    &  780          & $-$0.000488          \\
2,455,170.30081   & $\pm$0.00036    &  795          & $+$0.000095          \\
2,455,174.17830   & $\pm$0.00037    &  810          & $+$0.000055          \\
2,455,178.05523   & $\pm$0.00047    &  825          & $-$0.000563          \\
2,455,181.93463   & $\pm$0.00122    &  840          &                      \\
2,455,185.81061   & $\pm$0.00041    &  855          & $-$0.000348          \\
2,455,189.68853   & $\pm$0.00044    &  870          & $-$0.000059          \\
2,455,193.56649   & $\pm$0.00060    &  885          & $+$0.000225          \\
2,455,197.44446   & $\pm$0.00035    &  900          & $+$0.000457          \\
2,455,201.32484   & $\pm$0.00376    &  915          &                      \\
2,455,205.19999   & $\pm$0.00038    &  930          & $+$0.000226          \\
2,455,209.07744   & $\pm$0.00033    &  945          & $-$0.000408          \\
2,455,212.95618   & $\pm$0.00038    &  960          & $+$0.000118          \\
2,455,217.60977   & $\pm$0.00044    &  978          & $-$0.000166          \\
2,455,220.71211   & $\pm$0.00043    &  990          & $-$0.000332          \\
2,455,224.59094   & $\pm$0.00061    & 1005          & $+$0.000477          \\
2,455,228.46814   & $\pm$0.00070    & 1020          & $-$0.000236          \\
2,455,233.89723   & $\pm$0.00072    & 1041          & $-$0.000090          \\
2,455,236.22442   & $\pm$0.00032    & 1050          & $+$0.000447          \\
2,455,240.10160   & $\pm$0.00028    & 1065          & $-$0.000089          \\
2,455,243.97962   & $\pm$0.00046    & 1080          & $+$0.000255          \\
2,455,247.85651   & $\pm$0.00036    & 1095          & $-$0.000500          \\
2,455,251.73416   & $\pm$0.00031    & 1110          & $-$0.000468          \\
2,455,255.61230   & $\pm$0.00030    & 1125          & $+$0.000077          \\
2,455,259.48984   & $\pm$0.00037    & 1140          & $+$0.000040          \\
2,455,263.36772   & $\pm$0.00045    & 1155          & $+$0.000359          \\
2,455,267.24507   & $\pm$0.00046    & 1170          & $+$0.000162          \\
2,455,271.12260   & $\pm$0.00041    & 1185          & $+$0.000157          \\
2,455,275.00059   & $\pm$0.00046    & 1200          & $+$0.000622          \\
2,455,278.87777   & $\pm$0.00032    & 1215          & $+$0.000287          \\
2,455,282.75556   & $\pm$0.00036    & 1230          & $+$0.000570          \\
2,455,286.63250   & $\pm$0.00024    & 1245          & $+$0.000011          \\
2,455,290.50981   & $\pm$0.00027    & 1260          & $-$0.000173          \\
2,455,294.38740   & $\pm$0.00028    & 1275          & $-$0.000070          \\
2,455,298.26501   & $\pm$0.00026    & 1290          & $+$0.000057          \\
2,455,302.14214   & $\pm$0.00027    & 1305          & $-$0.000292          \\
2,455,306.01977   & $\pm$0.00044    & 1320          & $-$0.000137          \\
2,455,309.89818   & $\pm$0.00040    & 1335          & $+$0.000801          \\
2,455,313.77512   & $\pm$0.00029    & 1350          & $+$0.000272          \\
2,455,317.65206   & $\pm$0.00039    & 1365          & $-$0.000255          \\
2,455,321.52972   & $\pm$0.00038    & 1380          & $-$0.000061          \\
2,455,325.40743   & $\pm$0.00031    & 1395          & $+$0.000185          \\
2,455,329.28511   & $\pm$0.00025    & 1410          & $+$0.000401          \\
2,455,333.16213   & $\pm$0.00027    & 1425          & $-$0.000043          \\
2,455,337.81547   & $\pm$0.00040    & 1443          & $+$0.000339          \\
2,455,340.91746   & $\pm$0.00028    & 1455          & $+$0.000356          \\
2,455,344.79471   & $\pm$0.00028    & 1470          & $+$0.000139          \\
2,455,348.67209   & $\pm$0.00027    & 1485          & $+$0.000049          \\
2,455,352.54948   & $\pm$0.00032    & 1500          & $-$0.000035          \\
2,455,356.42625   & $\pm$0.00033    & 1515          & $-$0.000742          \\
2,455,360.30417   & $\pm$0.00029    & 1530          & $-$0.000305          \\
2,455,364.18158   & $\pm$0.00024    & 1545          & $-$0.000384          \\
2,455,368.05904   & $\pm$0.00037    & 1560          & $-$0.000421          \\
2,455,372.71243   & $\pm$0.00039    & 1578          & $-$0.000039          \\
2,455,375.81430   & $\pm$0.00031    & 1590          & $-$0.000184          \\
2,455,379.69239   & $\pm$0.00030    & 1605          & $+$0.000374          \\
2,455,383.56940   & $\pm$0.00026    & 1620          & $-$0.000165          \\
2,455,387.44712   & $\pm$0.00033    & 1635          & $-$0.000016          \\
2,455,391.32445   & $\pm$0.00048    & 1650          & $-$0.000284          \\
2,455,395.20258   & $\pm$0.00035    & 1665          & $+$0.000212          \\
2,455,399.07955   & $\pm$0.00049    & 1680          & $-$0.000498          \\
2,455,402.95845   & $\pm$0.00037    & 1695          & $+$0.000658          \\
2,455,406.83756   & $\pm$0.00215    & 1710          &                      \\
2,455,410.71354   & $\pm$0.00038    & 1725          & $-$0.000030          \\
2,455,414.59174   & $\pm$0.00033    & 1740          & $+$0.000075          \\
2,455,418.46982   & $\pm$0.00034    & 1755          & $-$0.000065          \\
2,455,422.34773   & $\pm$0.00030    & 1770          & $-$0.000387          \\
2,455,426.22680   & $\pm$0.00044    & 1785          & $+$0.000547          \\
2,455,430.10438   & $\pm$0.00062    & 1800          & $+$0.000115          \\
2,455,433.98185   & $\pm$0.00038    & 1815          & $-$0.000321          \\
2,455,437.86045   & $\pm$0.00039    & 1830          & $+$0.000454          \\
2,455,441.73775   & $\pm$0.00031    & 1845          & $-$0.000008          \\
2,455,445.61524   & $\pm$0.00029    & 1860          & $-$0.000231          \\
2,455,449.49316   & $\pm$0.00031    & 1875          & $+$0.000015          \\
2,455,453.37067   & $\pm$0.00030    & 1890          & $-$0.000117          \\
2,455,457.24815   & $\pm$0.00033    & 1905          & $-$0.000253          \\
2,455,457.24815   & $\pm$0.00033    & 1905          & $-$0.000253          \\
\enddata
\end{deluxetable}

\begin{deluxetable}{lcc}
\tablewidth{0pt}
\tablecaption{Parameters for the LTT orbits of KIC002856960.}
\tablehead{
\colhead{Parameter}        & \colhead{Value}        & \colhead{Unit}          
}                                                                          
\startdata                                                                 
$T_0$                      &  2,454,964.78879(14)   &   BJD                  \\
$P$                        &  0.25850790(12)        &   d                    \\
$a_{\rm b}\sin i_{3}$      &  0.453(41)             &   au                   \\
$\omega_{\rm b}$           &  353(4)                &   deg                  \\
$e_{\rm b}$                &  0.612(82)             &                        \\
$n_{\rm b}$                &  1.754(16)             &   deg d$^{-1}$         \\
$T_{\rm b}$                &  2,455,007.5(2.5)      &   BJD                  \\
$P_{\rm b}$                &  205.2(1.9)            &   d                    \\
$K_{\rm b}$                &  0.00208(19)           &   d                    \\
$f(M_{3})$                 &  0.295(26)             &   M$_\odot$            \\[1.0mm] \hline
$M_{3} \sin i_{3}$$\rm^a$  &  0.760(55)             &   M$_\odot$            \\
$a_{3} \sin i_{3}$$\rm^a$  &  0.274(10)             &   au                   \\
$\omega_{3}$               &  173(4)                &   deg                  \\
$e_3$                      &  0.612(82)             &                        \\
$P_3$                      &  205.2(1.9)            &   d                    \\
$\chi^2 _{\rm red}$        &  1.059                 &                        \\
\enddata
\tablenotetext{a}{M$_1$+M$_2$=0.46 M$_\odot$ is assumed. }
\end{deluxetable}

\end{document}